# Towards Lakosian Multilingual Software Design Principles

Damian M. Lyons[1], Saba B. Zahra[1], and Thomas M. Marshall[2]

[1]*Department of Computer and Information Science, Fordham University, New York, NY U.S.A.*
*{ dlyons, szahra }@fordham.edu*
[2]*Bloomberg LP, New York, NY U.S.A.*
*tmarshall2@bloomberg.net*

Abstract: Large software systems often comprise programs written in different programming languages. In the case when cross-language interoperability is accomplished with a Foreign Function Interface (FFI), for example pybind11, Boost.Python, Emscripten, PyV8, or JNI, among many others, common software engineering tools, such as call-graph analysis, are obstructed by the opacity of the FFI. This complicates debugging and fosters potential inefficiency and security problems. One contributing issue is that there is little rigorous software design advice for multilingual software.

   In this paper, we present our progress towards a more rigorous design approach to multilingual software. The approach is based on the existing approach to the design of large-scale C++ systems developed by Lakos. The FFI is an aspect of physical rather than logical architecture. The Lakosian approach is one of the few design methodologies to address physical design rather than just logical design. Using the MLSA toolkit developed in prior work for analysis of multilingual software, we focus in on one FFI – the pybind11 FFI.

   An extension to the Lakosian C++ design rules is proposed to address multilingual software that uses pybind11. Using a sample of 50 public GitHub repositories that use pybind11, we measure how many repositories would currently satisfy these rules. We conclude with a proposed generalization of the pybind11-based rules for any multilingual software using an FFI interface.

## 1 INTRODUCTION

Software systems, especially large ones, are often developed using more than one programming languages (Mushtak and Rasool 2015) (Mayer, Kirsch and Le 2017). These are often referred to as multilingual or polylingual (Barrett, Kaplan and Wileden 1996) software systems. Occasionally, multilingual software refers to software localization – customizing software to operate with a specific human language/culture; this is not its usage in this paper. In prior work (Lyons, Bogar and Baird 2018) we have discussed the historical and engineering trends that promote the development of multilingual software. We high-lighted a key challenge faced by developers and maintainers of multilingual software: "While it may be possible to automatically inspect individual language components of the codebase for software engineering metrics, it may be difficult or impossible to do this on a single accurate description of the complete multilingual codebase."

In a survey of software developers, Mayer et al. (Mayer, Kirsch and Le 2017) find that 90% request help in developing multilingual code bases. The issues they raise include redundancy between language functionalities, necessitating refactoring (Strien, Kratz and Lowe 2006); unexpected interactions between languages (Hong and al 2015), and security breaches in cross-language calls (Bravenboer, Dolstra and Visser 2010) (Lee, Doby and Ryu 2016). Moreover, it appears this issue will continue to grow in significance. In an analysis of over 3800 papers over a 15-year time period, Zaigham et al. (Zaigham, Rasool and Shehzad 2017) report that 23% of these papers were written in the two years prior to publication alone.

In our work, we have focused on the common *Foreign Function Interface* (FFI) style of cross-language interoperability. Examples include JNI, PyV8, pybind11, and many, many others. The advantages of examining this form of cross-language interoperability, apart from its popularity, include that it is amenable to static analysis of the source code by looking at a call-graph (since the cross-language transition is effected with a function call). However, the FFI is opaque: we cannot see what other-language functions are called within the FFI call. Thus, it becomes more difficult to verify software correctness, efficiency, and security. On top of this challenge, while an FFI package developer might (sometimes) make some recommendations as to how to safely use the interface, an application programmer is typically left to his or her own devices as to how to embed the FFI into the host language software. This can lead to a bewildering variety of designs for any one interface, and of course, a programmer might choose to use different interfaces for the same two languages at different places in the software to leverage ease of use or just plain familiarity. This lack of design guidance greatly complicates multilingual software engineering.

Grechanik et al. (Grechanik, Batory and Perry 2004) present a design approach for polylingual software. The problem they address is the $O(n^2)$ potential cross-language communications in $n$ programs. They propose a framework, called ROOF, that presents a single API for all cross-language communications. While this standardization clearly has benefits, we take a different approach: we believe software developers will use whatever FFI packages become available, so rather than try to change this behaviour, we limit ourselves to finding an approach that addresses the analysis of the 'messy' multilingual software thus created.

Lakos (Lakos 1996) addresses the problem of the physical design of large-scale C++ software. He proposes a set of guidelines for the physical architecture of large C++ codebases that has certainly stood the test of time (an extensively expanded version is scheduled by Addison-Wesley for 2021). It addresses issues such as how to structure source code files into components, packages, and package groups in a modular way, avoiding potentially troublesome, hidden linkages between the source code units that hinder efficiency, scalability and understandability. These troublesome linkage issues are similar to those caused by multilingual interoperability APIs and it is the thesis of this paper that some of the principles introduced for large-scale C++ physical design in (Lakos 1996), Lakosian principles, are relevant for multilingual design.

To illustrate this, we focus here on a common interoperability API, **pybind11** (Smirnoff 2017), which allows mutual calling between C++ and Python. While Python is often convenient for high level programming, C++ can offer a more efficient platform to implement numerical algorithms. Pybind11 was released in 2015 and has become a very popular alternative to the Boost.Python library (Abrahams and Grosse-Kunstleve 2003). To measure the popularity, we searched for public GitHub repositories using pybind11 and using Boost.Python. It has been 17 years since Boost.Python was released, and we found 478 repositories that referenced it. It has been only 4 years since pybind11 has been released and we found 233 repositories that reference it. The ratio of these time in existence to amount of repositories numbers would suggest that pybind11, at 58.2, is experiencing higher growth than the more mature Boost.Python, at 28.1. Public comments on programming forums typically cite pybind11's header-only nature, STL/Eigen support and full coverage of Boost.Python functionality as reasons for adoption. We take this step of picking one FFI to make our conclusions more concrete, and while we justify this choice for this paper, we could have selected a different one. We also specifically pick an FFI rather than a language such as Cython (http://cython.org) which aims to minimize the FFI aspect of calling C from Python; though given the popularity of Cython, we will come back to this later in the paper.

We begin in Section 2 by briefly reviewing those aspects of Lakosian design principles for large-scale C++ development that will prove useful. Multilingual software with the Foreign Function Interface (FFI) is introduced in Section 3 with pybind11 as a specific example. The MLSA (Multilingual Software Analysis) toolkit (Lyons, Bogar and Baird 2017) (Bogar, Lyons and Baird 2018), which will be used to investigate FFI usage in a collection of pybind11 repositories, is introduced in Section 4. Some common FFI practices will be identified that show analogous troublesome linkages to those first identified by Lakos in C++ physical design. As a solution, we will propose an extension of some Lakosian principles to address, first, pybind11/C++ repositories, and later (Section 6) FFI software in general. The pybind11 repository experimentation and results are reported in Section 5 and our conclusions presented in the final section.

## 2 LAKOSIAN DESIGN PRINCIPLES

Physical design is defined by Lakos as being concerned with the physical aspects of a software code repository: the directories, header files, source files, libraries and issues related to these. Contrast this with *logical design*, which relates to classes, functions and so forth. His motivation in considering physical design is the issue of scaling to very large software systems written in C++. Our motivation is different, relating as it does to the inter-language interface in multilingual software and issues relating to scaling, but we find that there are common concerns and solutions which can usefully be transferred from this prior work to our problem.

Lakos introduces some useful terminology and it is helpful to recapitulate it here:

- A *component* is a unit of physical design consisting of a source file (an implementation, a **.cpp** file for Lakos) and associated header file (an interface, a **.h** file for Lakos). All our work here will be component-based and we leave consideration of C++ modules for future work.
- A name is introduced into a name scope (namespace) via a *declaration* and is uniquely defined with a *definition.*
- Components are grouped into *packages* which are grouped into *package groups.*
- A *translation unit* is the union of all components being processed by a compiler at one time.
- A name has *external linkage* if it refers to the same object across multiple translation units.
- A name has *internal linkage* if it is only visible in its translation unit.

Among the many rules defined by Lakos to address issues in physical design, we identify a small subset that when interpreted in a multilingual context, as we will in a later section, are also important design rules. They are presented with their original intent here:

R1. The source file includes the header file as its first substantive line of code
R2. All logical constructs having external linkage defined in a source file are declared in the corresponding header file
R3. All constructs having external linkage declared in a header file are defined within the component.

While there are arguably additional Lakosian design rules that can address multilingual software, we select these rules because they are designed to address the problem of potential inconsistencies in name bindings in external linkages. They ensure that externally linked functions and objects are defined in a predictable, easy to find, and easily understood manner.

## 3 MULTILINGUAL SOFTWARE

Modern software development is increasingly multilingual: Developers might build a software project from different language components for functionality or style reasons. In prior work (Lyons, Bogar and Baird 2017) (Bogar, Lyons and Baird 2018) (Lyons, Bogar and Baird 2018), we have looked at the problem of providing tools for analyzing multilingual code bases to address software engineering and security concerns. We proposed a software architecture for lightweight multilingual software analysis tools – the MLSA architecture and introduced some tools for generating multilingual call-graphs, which we argued can be the basis for the addressing many useful software engineering and security concerns.

### 3.1 Foreign Function Interface

Programs in one language can communicate with software written in a second language through several means (Grimmer, et al. 2018): message passing, foreign-function interface, and multi-language run-time. Rather than proposing any new scheme for writing, including or executing multilingual software in this paper, we have instead focused on processing existing and widely used FFI packages, e.g., JNI (C with Java), *Python.h* (C/C++ with Python), PyV8 (JavaScript with Python), Emscripten (JavaScript from C++), JQuery (Python from JavaScript), and so forth. As we initially stated, advantages of focusing on this form of cross-language are that it is quite common and it is amenable to static analysis of the source code by looking at a call-graph. Message passing is more difficult to analyze statically (Bronevetsky 2009), and a common language runtime analysis must be done on the common language bytecode.

All of these FFI packages differ in their syntax and use and there are no standards or guidelines that specifically address their use. In fact new cross-language interfaces are defined all the time: Boost.Python (Abrahams and Grosse-Kunstleve 2003) introduced a powerful library for binding C++ classes and functions to Python, a great improvement over the interoperability provided by Python's C/C++ interface (Python 2.7 documentation 2010). Subsequently, pybind11 was developed as a light-weight, header-only interface (Smirnoff 2017) with similar functionality. For historical reasons, all three are still used, and it is not unlikely that new ones will be developed!

All these interoperability interfaces provide the common function of mapping from the namespace of one programming language to that of another. But in the Lakosian vocabulary, they map from the namespace of one translation unit to that of another. In that framework, every translation unit is C++ code, Fig. 1(a) whereas we propose to relax that constraint. Hence, FFIs define an external linkage between the two language translation units, Fig.1(b) – a topic about which Lakos has design advice that we can leverage. Our approach to addressing this issue with

Lakosian design principles will be to work from the specific to general, starting with a pybind11 example.

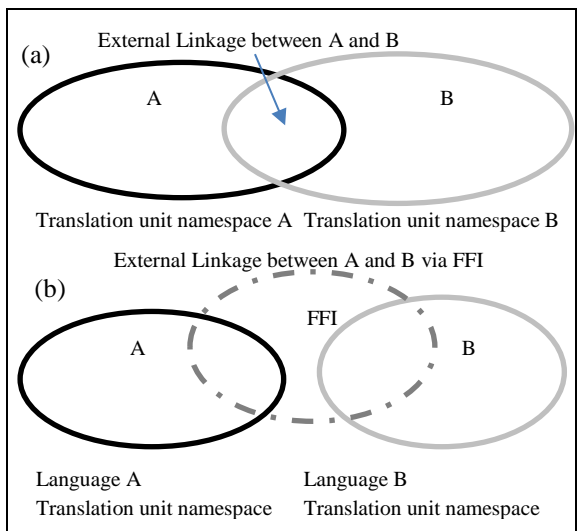

**Figure 1:** (a) External linkage between C++ translation units; (b) Foreign-function interface between multilingual translation units.

### 3.2 Standard pybind11 case

A straightforward example of the use of pybind11 to call a C++ function from a Python script `A.py` is shown in Fig. 2(a). A C++ file `B.cpp`, shown in Fig. 2(b), defines a function `f`, and establishes that it can be called in Python (with the same name `f` in this example). After compilation, a binary file is generated that can be imported as a Python module.

| `#Python file A`<br><br>`# C++ functions`<br>`import B`<br>`.`<br>`.`<br>`# call C++ function`<br>`x = B.f(34)`<br>`.`<br>`.`<br>`.`<br>(a) | `//C++ file B`<br><br>`#include <pybind11>`<br><br>`// declaration`<br>`int f(int a);`<br><br><br>`// bind python "f" to f`<br>`PYBIND11_MODULE(B,m){`<br>`m.def("f",&f)`<br>`}`<br>(b) |

**Figure 2**: pybind11 example.

Let us begin by considering the software in Fig. 2 as two components. Each component has an interface (header file) and implementation (source file). This separation is important in physical design. The lines shown in bold font in Figs. 2(a) and (b) is the header file information and those below it, the source file information.

To respect R1-3 of Section 2, it necessary that the header file `B.h` contains (at least) the material in bold in Figure 2(b) and the source file `B.cpp` contains (at least) the remainder. The rationale is a little different from the regular C++ case but reflects the same concerns with external linkage, or name binding. The Python file `A.py` needs to import declaration information for the external linkages (C++ names) it uses (R2). For this example, that information is in the `m.def` commands in `B.cpp` and **not in** `B.h`, which has the declarations of the C++ functions referenced by these bindings. Thus, the implementation declarations should be separate from the bindings, which should be in a source file named for the module. (An informal rule advocated by the pybind11 developers). The implementation declarations should be included in the header file for the component, and definitions in a separate source file, providing insulation for the module implementation.

Summarizing, and based on R1-R3 in Section 2:

M1. The python component imports the pybind11 generated module by name, in standard python syntax.
M2. The source file of the binding component must be named after the module, so that there is a transparent connection to the interface with the important statement (R2).
M3. The header file of the binding component should declare the implementation of the bound functions (R3).

In addition to following Lakosian design principles for physical architecture, or perhaps because of it, these rules also simplify the static analysis of multilingual code bases by making it readily clear in what files functions are located. The multilingual software analysis toolkit MLSA is described in the next section as prelude to using the toolkit. Public GitHub repositories using pybind11 will be analysed with MLSA to determine how many abide by the proposed principles and in what ways they differ.

### 4 THE MLSA TOOLKIT

Developers may build a software project (here, the *input data* for our MLSA analysis tools) from different language components for functionality or style reasons, while at the same time companies maintain a commitment to language already used in company software, leading to an increasingly crowded and complex landscape of multilingual software development. Any solution that is narrowly focused on the existing state-of-the-art will find itself quickly outdated as new languages or interoperability APIs or language embeddings appear. For this reason, we proposed the following design principles for a software architecture for MLSA (Lyons, Bogar and Baird 2017) (Fig. 3).

1. **Lightweight**: Computation is carried out by small programs, which we call *filters*, that communicate results with each other. A specific software analysis is built as a *pipeline* of these programs.
2. **Modular**: Filter programs accept input and produce output in a standard form, to allow modules to be substituted or added with minimum collateral damage.
3. **Open**: MLSA uses open-source software for monolingual processing and for display.
4. **Static Analysis**: MLSA uses static source code analysis as its principle approach

Illustrating the lightweight design principle, each filter is a small standalone program, implementing a single analysis. The only constraint on the filter is on the format of its input and output. Filter programs can be written in different languages as shell scripts.

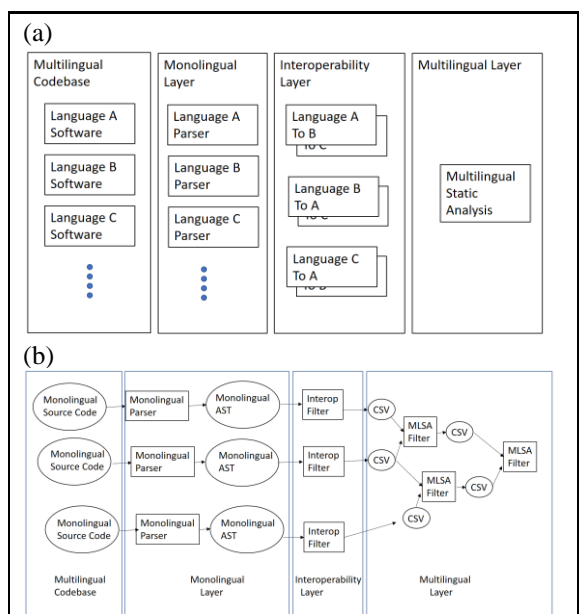

**Figure 3:** MLSA Software Architecture (a); example filter pipeline, from (Lyons, Bogar and Baird 2017) (b).

The modularity design principle means that extending MLSA to handle a new cross-language interface is as simple as extending the parts of the processing pipeline with scripts designed for the new interface.

### 4.1 Multilingual call-graph analysis

We have argued in prior work (Lyons, Bogar and Baird 2017) (Lyons, Bogar and Baird 2018) (Bogar, Lyons and Baird 2018) that call-graph analysis is one productive tool for investigating software engineering properties of multilingual code, and we have focused our research on what is required to construct a multilingual call-graph. In addition to the challenges of monolingual call-graph construction, e.g., (Ali and Lhotak 2012) (Bacon and Sweeney 1996) (Bogar, Lyons and Baird 2018), multilingual call graphs have to include edges that span one language to a second. Since it is very typical in an FFI for the name of the function in the second language to be specified by the value of an argument to a function, and not as a name or label in the host name space, static analysis of the code becomes challenging. In (Lyons, Bogar and Baird 2018), we propose a solution to this problem based on a Reaching Definitions Algorithm (RDA) (Nielson, Nielson and Hankin 2005) to statically resolve argument values. Using this, we demonstrated multilingual call graphs for software code bases using C/C++, Python and JavaScript languages. We showed how the call graph could be used to identify potential software security hazards and other software engineering concerns that would normally be hidden by the opacity of the cross-language interface.

### 4.2 MLSA pybind11 filter

To include the pybind11 cross-language interoperability interface to the repertoire of FFI that MLSA handles, it was only necessary to build an additional MLSA filter and add it to the multilingual call-graph pipeline in Fig. 2 of (Lyons, Bogar and Baird 2018). That pipeline is summarized below, for convenience. For simplicity, and since it suited the goals of this paper, only the Python calling C++ direction was implemented (our review of GitHub repositories shows this was by far more common), and only the function call interface (`.def`) was implemented and not the variable (`.attr`) interface.

The MLSA multilingual call-graph pipeline (Lyons, Bogar and Baird 2018) proceeds as follows:
1. MLSA begins by generating Abstract Syntax Trees (ASTs) for each monolingual program.
2. From these, it extracts all the function definitions and calls, and writes them to csv-formatted data files. (Bogar, Lyons and Baird 2018)
3. Filter programs for various cross-language interoperability FFIs post-process these files, replacing the cross-language FFI call with the cross-language function being called (in so much as that can be statically determined by RDA).
4. The resultant data files are then merged to produce a single multilingual call graph.

The MLSA pybind11 filter is just another filter added to stage 3 of this processing pipeline. The filter searches the datafiles generated from the Python AST looking for a function call in a module that has been defined in C++ rather than Python. (It knows this from inspecting the data files generated in stage 2.) The datafile for the C++ module is then scanned for references to the `.def` function. The arguments to `.def` are processed to determine the name of the target function. Only literal values, or variables previously assigned literal values (found by RDA analysis) are currently allowed. The python function call is replaced in the datafile by the C++ function.

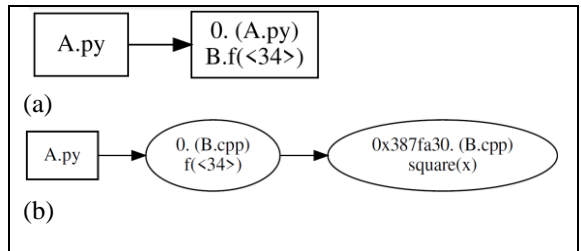

**Figure 4:** Multilingual call-graph extracted by MLSA from the example in Fig. 2: (a) just A.py and (b) for A.py and B.cpp/.h *Node shape: Rect: Python; Ell: C/C++.*

Fig. 4 shows the multilingual call graph generated in this way from the example in Fig. 2. Fig. 4(a) shows the call graph for just the Python component `A.py`. Fig. 4(b) shows the call graph after the pybind11 filter has acted. The python function call `B.f()` has been correctly bound to the C++ function `f()` and its implementation subgraph included. Different languages are indicated in MLSA call graphs with different node shapes: here, rectangular indicates Python and ellipse indicates C++. Note that the implementation of `f()` is not shown in Fig.2 but is shown here: A call from the C++ `f()` function to the C++ `square()` function.

The MLSA pybind11 filter provides a convenient lens to study how pybind11 is used in practice, and to evaluate to what extent the proposed Lakosian inspired design rules are followed. In the next section we present the results of our experimental analysis of public GitHub repositories.

## 5 EXPERIMENTAL RESULTS

GitHub has become a popular platform for collaborative software development, with over 30M users and 100M repositories (github.com/about). To measure how pybind11 is used by software developers, and how often our proposed design rules reflect common practice, we collected public GitHub repositories that use pybind11. A simple keyword search showed over 900K repositories using Python and almost 500K repositories using C++. Of these almost 12K used both C++ and Python. However, only 233 specifically mentioned pybind11, and we restricted our attention to these.

Many of the repositories just contained examples or tutorials for pybind11 and in general we rejected all but one copy of these small repositories until we had reached 50 samples – an arbitrary limit selected to yield generalizable results. Table 1 describes the sizes of the repositories.

| Table 1: Details of Sample Repositories. | |
|---|---|
| Num. repositories | 50 |
| Num. pybind11 modules | 449 |
| Total num. files | 3712 |
| Num. PY files | 540 |
| Num. CPP files | 748 |
| Total lines of code | 4.2M |
| Max, min, avg. num. files | 1131, 3, 75 |
| Max, min, avg. lines of code | 2.3M, 312, 83K |

The MLSA pybind11 filter was augmented to count how often the repositories complied with our extended Lakosian design rules. The results are in Table 2.

| Table 2: Summary Results of Study. | | |
|---|---|---|
| | Rep. | Mod. |
| Num. rep./mod. meeting M1-M3 | 8 (16%) | 360 (80%) |
| Binding misname (fails M2) | 27 (54%) | 40 (9%) |
| Impl. in binding (fails M3) | 24 (48%) | 53 (12%) |

Table 2 reports that 16% of the repositories strictly followed the guidelines proposed. That means *all* 360 modules in those repositories followed the rules. The small repository number and large module number indicate that the repositories following the rules were in fact all the largest repositories in the collection. We contend that this large minority figure could be considered a vote in favour of our proposed guidelines. Given the variety of usage observed, we summarized failure to follow the guidelines into two categories: Where the binding component was not named after the module (fails M2; 54%) and where the implementation was included alongside the binding (fails M3; 48%). A repository was counted as failing M2 or M3 if *any* module in the repository failed. As can be seen from the smaller module percentages for this, many repositories contained a mix of modules meeting and not meeting each of the two criteria. Based on this study, 84% of the repositories have potential software engineering challenges related to ambiguity in how they defined external linkages. Several recurring themes were identified in non-compliant repositories; we discuss these in more depth below.

<u>Anonymous functions.</u> We consider any function, such as a lambda function, which is defined directly in the binding as an anonymous function. This approach fails to meet design rule M3. An example is shown in Fig. 5. While lambda functions are elegant solutions to certain problems, anonymous functions of any kind are more difficult to track. Our approach in MLSA is to flag these for closer (manual) inspection when detected. We observed that 20% of the repositories used lambda functions in *any* module. Only 4% of modules used lambda functions, however, approximately half of which were used for member variable get/set functionality.

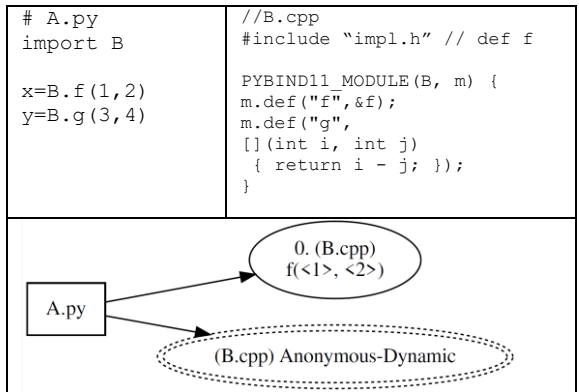

**Figure 5:** Anonymous function example.

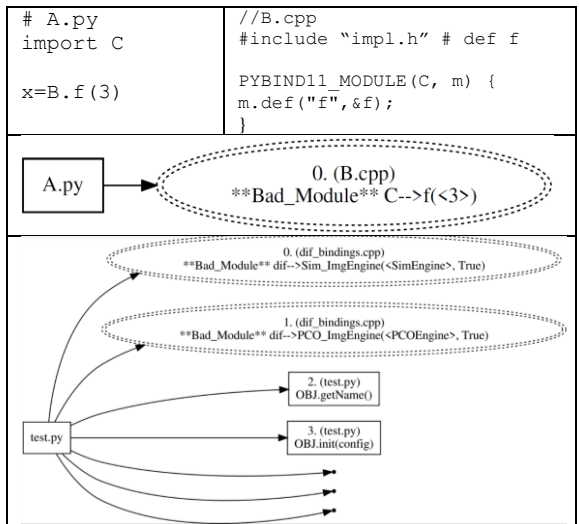

**Figure 6:** Misnamed module example (top two rows, explanatory example; bottom row, summarized example call-graph from repository collection).

Misnamed module. Included in this category is anything that puts the binding information in a place that is not clearly visible by inspection of the software or static analysis. The obvious example (shown in Fig. 6) is where the binding component is not named with the module name. MLSA can still find the binding but it flags the situation for correction, since the software becomes more difficult to maintain when the external linkages are not readily visible in a single location. Other cases of this include multiple modules in one binding (makes clear naming difficult), or binding information in a different location from the location the module will be imported from (a directory named for the module, for example). We found **no example** where there were multiple modules defined in a single binding.

# 6 PROPOSED LAKOSIAN ML DESIGN RULES

This paper has focused on pybind11 as a specific example for developing multilingual software design guidelines. We generalize our Lakosian inspired physical design rules M1-M3 for any FFI from a *host* language to a *target* language.

ML1. The host component imports/includes a single *uniquely named FFI interface file*.
ML2. The source file of the FFI binding component in the target language must be *named after the FFI interface file*, so that there is a transparent connection to the interface (R2).
ML3. The source file of the binding component should *import, or include* declarations of, the implementation of the bound functions (R3) in the target language.

While we argue that this extension can be justified based on our pybind11 examples and the original Lakosian rules, a realistic validation requires it should be evaluated against a range of FFI. That work is beyond the scope of this paper in general, but we will look at just one important case: How would this specialize to Cython?

Cython is a static compiler that accepts a superset of Python allowing a programmer to mix C/C++ declarations and definitions in Python-like code which is then translated to C for efficient implementation. In particular, programmers can write modules that can be imported and called from Python, in a manner analogous to pybind11. Our general ML design rules can be applied to this usage of Cython as follows:

C1. The python component imports the Cython generated module (created by the build file) by name, in standard python syntax.
C2. The binding file (~.pyx) should be named the same as the binding module name in the build file (~setup.py).
C3. The binding file should include a header file containing declarations of the C functions used in the binding functions in (~.pyx).

# 7 CONCLUSIONS

Large software systems often have components written in different languages. The FFI approach to cross-language interoperability is popular and relatively easy to use, mimicking as it does a host language function call. Our approach is to accept that programmers will continue to use existing and new FFI interfaces, and we do not try to present a new multilingual interface concept or formalism. Instead, we have developed a lightweight toolkit, MLSA (Lyons, Bogar and Baird 2018) for analysing such multilingual systems and identifying software engineering and security issues. In particular, given the nature of the FFI, we have focused on call-graph analysis (Bogar, Lyons and Baird 2018). In this paper, we have addressed the problem that there is little in the way of design assistance for FFI based multilingual software, comparable to the Lakosian design method of (Lakos 1996) for C++.

We have proposed an extended application of the Lakosian design method to multilingual software. In particular, we applied this to software written using the popular pybind11 FFI to call C++ from Python.

We reported on our study of 50 public GitHub pybind11 repositories: While 16% already adhered to our guidelines, 84% did not. Of the non-compliant cases, binding file naming and use of lambda functions are the most often issues. Based on our experience with pybind11, we proposed a more general set of multilingual design rules. In future work, we plan to evaluate these rules against other FFI. Finally, we believe we have just begun the task of understanding how Lakosian physical design rules hold value for the design of multilingual software.

## ACKNOWLEDGEMENTS

Lyons and Zahra are partially supported by grant DL-47359-15016 from Bloomberg L.P.